# A non-energetic mechanism for glycine formation in the interstellar medium

**S. Ioppolo,[1*] G. Fedoseev,[2,3] K.-J. Chuang,[4] H. M. Cuppen,[5] A. R. Clements,[6] M. Jin,[6] R. T. Garrod,[6,7] D. Qasim,[2] V. Kofman,[2†] E. F. van Dishoeck,[8] H. Linnartz[2]**

[1]School of Electronic Engineering and Computer Science, Queen Mary University of London, London, UK
[2]Laboratory for Astrophysics, Leiden Observatory, Leiden University, Leiden, the Netherlands
[3]Research Laboratory for Astrochemistry, Ural Federal University, Ekaterinburg, Russia
[4]Laboratory Astrophysics Group of the Max Planck Institute for Astronomy, Institute of Solid State Physics, Friedrick Schiller University Jena, Jena, Germany
[5]Theoretical & Computational Chemistry, Institute for Molecules and Materials, Radboud University, Nijmegen, the Netherlands
[6]Department of Chemistry, University of Virginia, Charlottesville, VA, USA
[7]Department of Astronomy, University of Virginia, Charlottesville, VA, USA
[8]Leiden Observatory, Leiden University, Leiden, the Netherlands
**† Current address:** NASA Goddard Space Flight Center, Greenbelt, MD, USA
*** E-mail:** s.ioppolo@qmul.ac.uk

## ABSTRACT

The detection of the amino acid glycine and its amine precursor methylamine on the comet 67P/Churyumov-Gerasimenko by the Rosetta mission provides strong evidence for a cosmic origin of prebiotics on Earth. How and when such complex organic molecules form along the process of star- and planet-formation remains debated. We report the first laboratory detection of glycine formed in the solid phase through atom and radical-radical addition surface reactions under cold dense interstellar cloud conditions. Our experiments, supported by astrochemical models, suggest that glycine forms without the need for 'energetic' irradiation, such as UV photons and cosmic rays, in interstellar water-rich ices, where it remains preserved, in a much earlier star-formation stage than previously assumed. We also confirm that solid methylamine is an important side-reaction product. A prestellar formation of glycine on ice grains provides the basis for a complex and ubiquitous prebiotic chemistry in space enriching the chemical content of planet-forming material.

## Introduction

The unravelling of the formation and distribution of complex organic molecules (COMs) in space is pivotal to our understanding of the initial conditions for the emergence of life on Earth. Glycine ($NH_2CH_2COOH$), the simplest amino acid, has been detected together with methylamine ($NH_2CH_3$), one of its possible precursors, and other organic compounds in the coma of comets such as Wild 2 by the Stardust mission and 67P/Churyumov-Gerasimenko (67P) by the Rosetta mission[1,2]. In recent years, tens of indigenous amino acids have also been found in meteorites[3,4]. While the formation path of amino acids in carbonaceous chondrites has been suggested to occur mainly through aqueous alterations in deep layers under the surface of planetesimals of the Solar System[5], it has been recently confirmed that the glycine found embedded in sublimating water ice from dust particles that were ejected from the nucleus of comet 67P is consistent with pristine material that has not been significantly altered either by heat or liquid water[6]. Consequently, there is a strong evidence that comets are the most primitive planetary bodies in our Solar System and that the building blocks of life present in their ices have an interstellar origin[7,8].

Current laboratory and modelling work have so far indicated that the interstellar formation of glycine, alanine, serine and other prebiotic species occurs by means of 'energetic', e.g. UV photon, cosmic ray, electron, X-ray, thermal, processing of interstellar ices. It is accepted that UV photon and ion irradiation at 10-20 K, as well as radical-radical recombination reactions across a temperature range ~40-120 K simulating the warm-up phase of hot cores, can both lead to the formation of amino acids in interstellar ice analogues in *later* stages of star formation[9-15]. However, the same laboratory experiments also show that the 'energetic' processes that induce formation of amino acids in space cause their chemical alteration and destruction at higher irradiation doses[16,17]. Therefore 'non-energetic' formation routes, e.g. atom-addition surface reactions in absence of any 'energetic' trigger, should also be investigated, specifically as these processes govern the chemistry in dark and dense clouds, i.e. during the *early* stages of star formation. Moreover, water ($H_2O$) ice, a polar environment, has been found to be the most favorable place where primeval glycine and possibly other prebiotic species can survive throughout the evolution of star and planet formation[18]. The recent detection of glycine embedded in sublimating water ice from comet 67P agrees with the picture that glycine is formed and preserved in a water-rich ice environment in the interstellar medium (ISM)[6]. Nevertheless, to date there is not a direct prove that glycine is present in the ISM. Extensive observational searches for glycine gas-phase signatures in the ISM have not resulted in its unambiguous identification so far[19], but several possible glycine precursors, such as methylamine, have been already detected across a variety of star formation environments, e.g. toward Sgr B2 and hot cores associated with the high-mass star-forming region NGC 6334I, confirming the ubiquity of such species [20]. Furthermore, although to date only simpler ice species have been unambiguously observed, it is expected that the spectral sensitivity of the upcoming James Webb Space Telescope (JWST) mission will increase chances to observe larger, prebiotic species, possibly also including amino acids, in the ISM in the solid state across a variety of astronomical cold environments[21].

Pivotal to a future detection of glycine in the ISM is our deeper understanding of its formation pathway and distribution beyond the Solar System. For instance, recent carbon atom-addition experiments in liquid helium droplets suggested that glycine may be already available in interstellar regions where atomic carbon is abundantly present, e.g. edges of molecular clouds[22]. Moreover, several COMs such as glycolaldehyde, ethylene glycol and glycerol have been proven to form efficiently in the solid phase through 'non-energetic' atom and radical-radical addition reactions under conditions relevant to interstellar dark clouds[23-26]. This work is an extensive joint laboratory and astrochemical modeling effort that strongly supports the relevance of a 'non-energetic' primeval origin of glycine in interstellar ices. Here experiments are performed under relevant prestellar conditions, i.e. on a low temperature (13-14 K) ice surface. The ultimate aim of our work is to further characterize the degree of chemical complexity that can be reached in interstellar ice analogues through a network of 'non-energetic' surface reactions at early stages of star formation.

## Results and Discussion

The laboratory formation of glycine presented here follows a systematic and extensive study of the many individual species and radicals involved in the 'non-energetic' surface reaction scheme depicted in Figure 1. Our experiments aim to ultimately simulate interstellar relevant surface reaction routes tested under fully controlled laboratory conditions. Here we demonstrate that interstellar glycine forms in the first water-rich ice layer covering bare dust grains together with more abundant simple ice species such as carbon dioxide ($CO_2$), ammonia ($NH_3$) and methane ($CH_4$). In space, most of the CO present in a polar environment is converted to $CO_2$ through the surface reaction CO + OH[27]. Hydroxy radicals (OH) are efficiently formed on the surface of interstellar ice grains during the formation of a water layer through atom addition surface reactions[28,29]. During surface ammonia and methane formation[30,31], it is likely that methylamine and its precursor $NH_2CH_2$ radical are also produced in the same ice layer through radical-radical recombination reactions and can then take part in more complex surface reaction routes leading to the formation of glycine and other amino acids[32].

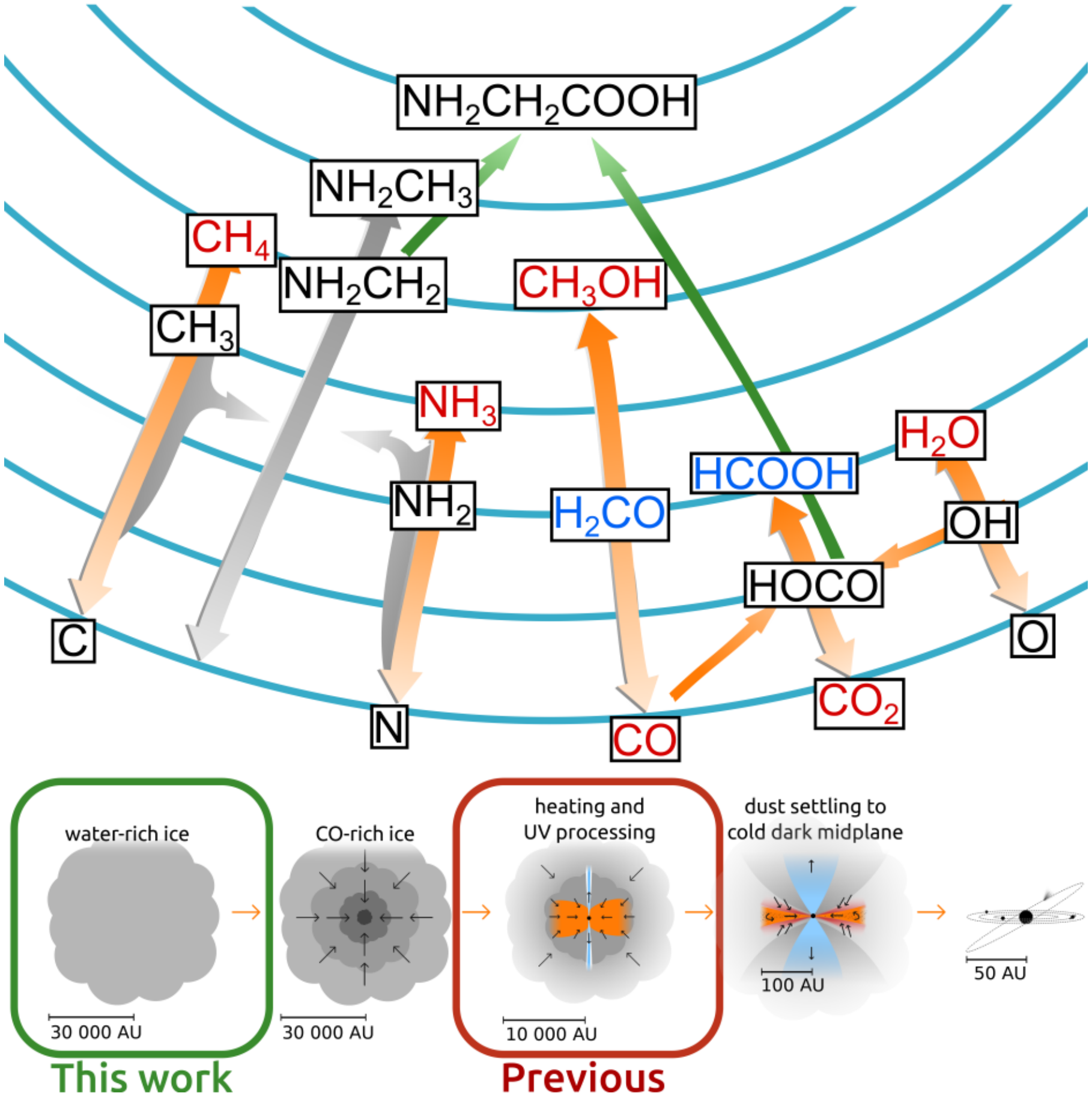

**Figure 1.** Schematic of surface reaction routes leading to the formation of glycine in a water-rich ice during early stages of low-mass stellar evolution. Top: The reaction scheme depicted above is based on results from this work, ref.[31] and references reviewed in ref.[29]. Each blue ring in the figure represents an additional hydrogenation step. Therefore, hydrogenation-addition reactions are depicted with upward arrows; downward arrows indicate hydrogenation-abstraction reactions; and horizontal (or diagonal) arrows are radical-radical recombination reactions. Orange arrows depict 'non-energetic' surface reaction routes tested under laboratory conditions. Green arrows are reactions investigated in this work. Gray arrows are surface reactions not investigated in literature. In a water-rich ice, H and OH addition/abstraction reactions result in stable species, like $H_2O$, $CH_4$, $NH_3$ and $CO_2$, and intermediate radicals, such as $CH_3$ and $NH_2$. Further radical-radical recombination reactions allow for the formation of methylamine, $NH_2CH_3$. Glycine (top box) is finally formed through the recombination of the HO-CO complex and the $NH_2CH_2$ radical. Species which are unambiguously detected in ices in prestellar cores are depicted in red; tentatively detected species are in blue. The remainder of the species involved in this reaction scheme are in black. Bottom: A schematic of the current view of a stellar evolution adapted from ref.[49]. The green box indicates the prestellar stage where glycine is formed on water-rich ice grains according to this work. Highlighted in a red box is the later evolutionary stage where glycine was previously thought to exclusively form through 'energetic' processing of interstellar evolved ices[7-13].

**Table 1.** List of the most relevant glycine formation and deposition experiments.

| # | Glycine Formation | T K | Flux (CO) $10^{12}$ mol. $cm^{-2}$ $s^{-1}$ | Flux ($NH_2CH_3$) $10^{12}$ mol. $cm^{-2}$ $s^{-1}$ | Flux ($O_2$) $10^{12}$ mol. $cm^{-2}$ $s^{-1}$ | Flux (H) $10^{12}$ mol. $cm^{-2}$ $s^{-1}$ | time min | TPD K/min |
|---|---|---|---|---|---|---|---|---|
| 1 | $CO + NH_2CH_3 + O_2 + H$ | 13 | 1.3 | 0.4 | 0.7 | 16 | 360 | 5 |
| 2 | $CO + NH_2CH_3 + {}^{18}O_2 + H$ | 14 | 1.3 | 0.4 | 0.7 | 16 | 360 | 5 |
| 3 | ${}^{13}C^{18}O + NH_2CH_3 + O_2 + H$ | 13 | 1.3 | 0.4 | 0.7 | 16 | 360 | 5 |
| 4 | ${}^{13}C^{18}O + NH_2CH_3 + {}^{18}O_2 + H$ | 14 | 1.3 | 0.4 | 0.7 | 16 | 360 | 5 |
| 5 | $CO + NH_2CH_3 + O_2 + D$ | 14 | 1.3 | 0.4 | 0.7 | 16 | 360 | 5 |
| 6 | $CO + NH_2CH_3 + O_2 + H$ | 13 | 1.3 | 0.4 | 0.7 | 16 | 360 | 10 |
| | **Glycine Deposition** | **T** K | **Flux** $10^{12}$ mol. $cm^{-2}$ $s^{-1}$ | | | | **time** min | **TPD** K/min |
| | $NH_2CH_2COOH$ | 12 | 16 | | | | 20 | 5[a] |

[a] TPD was performed in steps to acquire IR spectra at different temperatures.

To study the formation of glycine in the laboratory, we simultaneously deposit (i.e. co-deposit) a glycine precursor, $NH_2CH_3$, with CO molecules and have OH radicals formed in the ice at 13 K. In our experiments, all radicals are formed in the ice via 'non-energetic' H- and OH-induced addition/abstraction surface reactions involving a deposited stable molecule[33]. For instance, to form hydroxy radicals (OH) we co-deposit molecular oxygen ($O_2$) and hydrogen atoms (H) onto our cold substrate. The following reaction chain leads to the formation of OH radicals: $O_2 + H \rightarrow HO_2$, $HO_2 + H \rightarrow 2OH$. Newly formed OH can then react both with CO molecules to form the HO-CO complex and with $NH_2CH_3$ to form the $NH_2CH_2$ radical through the abstraction of a hydrogen atom. During the same process, H atoms can also abstract a hydrogen from the methyl group of $NH_2CH_3$, increasing the amount of $NH_2CH_2$ radicals in the ice (see Supplementary Information). The further barrierless recombination of HO-CO and $NH_2CH_2$ leads to the formation of glycine at 13 K. In this reaction scheme, $CO_2$ efficiently forms through the direct dissociation or hydrogenation of the HO-CO complex, which are competing processes that reduce the amount of available HO-CO in the ice[34]. Therefore to improve the detection limit of glycine, we selected relative co-deposition ratios for our initial species that would optimize the formation of OH radicals in the ice ($CO:NH_2CH_3:O_2:H$=3:1:2:40; for more details see Table 1). Under such conditions, secondary products, e.g. water and hydrogen peroxide ($H_2O_2$), are expected to be efficiently formed[35] together with traces of formaldehyde ($H_2CO$) and methanol ($CH_3OH$) through the hydrogenation of CO[36]. Thus, although the initial ice mixture composition may not be representative of an interstellar ice layer, the final ice obtained upon surface reactions is analogue to a water-rich interstellar ice environment.

As discussed in the Methods, we observe all species *in situ* at low temperatures by using Reflection Absorption Infrared Spectroscopy (RAIRS), i.e. without external manipulation and possible contamination of the formed molecules, and by means of Quadrupole Mass Spectrometry (QMS) upon heating the ice at a constant rate up to its complete desorption, a method known as temperature programmed desorption (TPD). QMS-TPD data show that in our experiments glycine desorbs at 245 K and is detected from its mass signals 30 and 75 m/z (top-left panel of Figure 2). According to the NIST mass-database, the main fragmentation mass of glycine upon electron impact ionization is 30 m/z (i.e. $NH_2CH_2$), while 75 m/z (i.e. non-dissociated glycine) is here chosen because it does not overlap with any fragmentation pattern of other species co-deposited in the ice. Detection of glycine is further confirmed by experiments which start with isotopolog precursors to form $NH_2CH_2CO^{18}OH$, $NH_2CH_2{}^{13}C^{18}OOH$ and $NH_2CH_2{}^{13}C^{18}O^{18}OH$ (see Figure 2 showing experiments 1-4 from Table 1). The mass signals presented here highlight the expected mass shifts for glycine when isotopologs are used confirming its detection (see left panel of Figure 3). In our experiments 1-4 of Table 1, mass 30 m/z is also the main fragmentation mass for methylamine that partially desorbs at 110 K but also can co-desorb with water and hydrogen peroxide at temperatures below 200 K[37]. Because of the relatively fast heating rate (5-10 K/min) of the TPD experiments and the relatively low pumping speed of the main-chamber turbo pump, a non-negligible contribution to mass signal 30 m/z at 245 K is still due to residual methylamine present in the chamber in the gas phase. Nevertheless, in all panels of Figure 2 a weak desorption peak for mass signal 30 m/z is visible at 245 K on top of the decreasing background signal. Right panels of Figure 3 confirm that all newly formed glycine isotopologs have similar fragmentation patterns for mass signals 30 and 75 m/z that are in good agreement with the NIST mass-database (orange histogram). Additionally, a series of control experiments presented in detail in Supplementary Information are performed to ensure the unambiguous detection of glycine during QMS-TPD experiments.

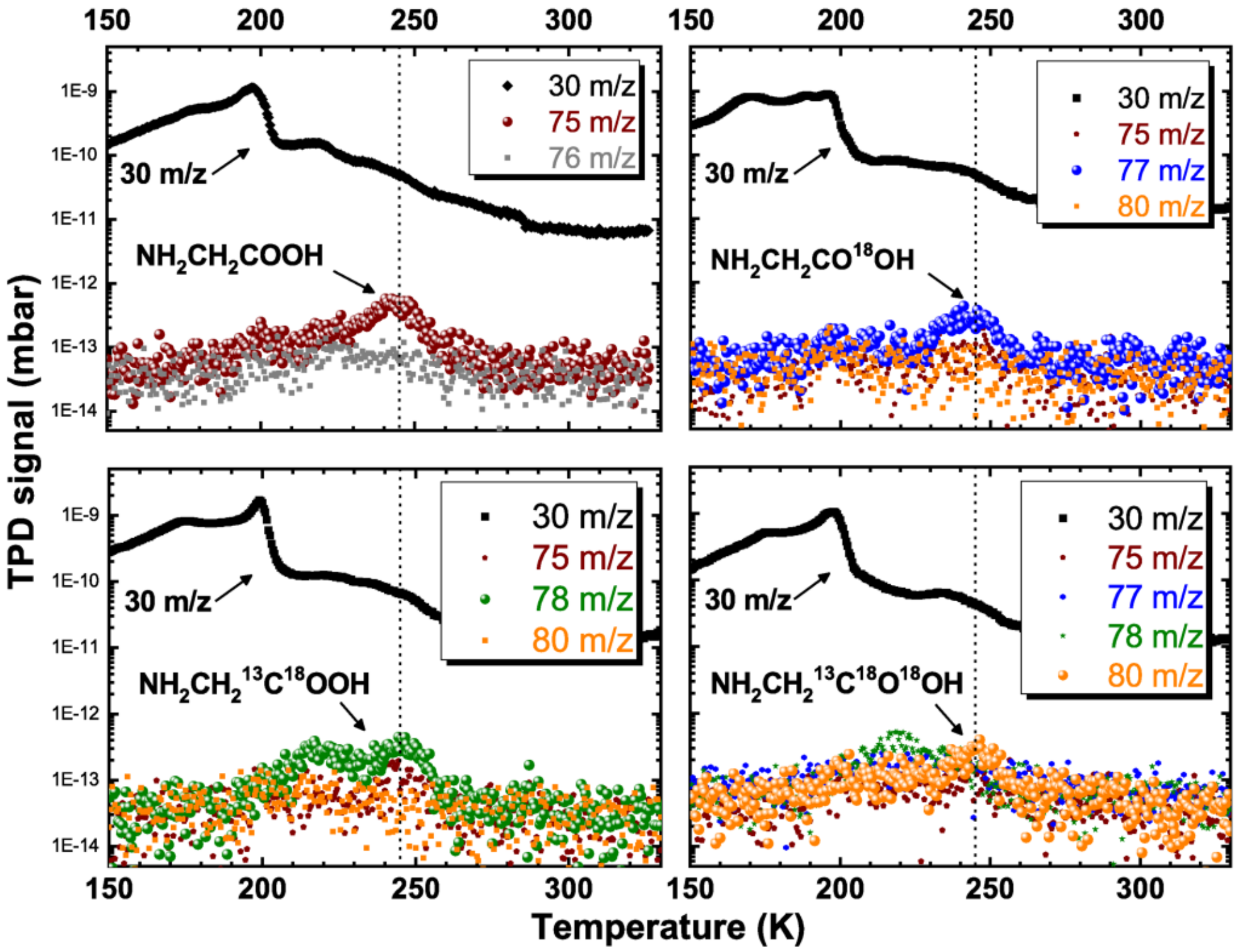

**Figure 2.** Quadrupole Mass Spectrometer-Temperature Programmed Desorption (QMS-TPD) data of four equivalent laboratory experiments on the surface formation of glycine and its isotopologs. In all panels, mass signal 30 m/z, i.e. the main contribution of fragmented glycine according to the NIST database, is depicted. Mass signals 75, 77, 78, and 80 m/z represent the signal from the non-fragmented isotopologs of glycine from experiments 1-4 of Table 1, i.e. $NH_2CH_2COOH$ (top-left), $NH_2CH_2CO^{18}OH$ (top-right), $NH_2CH_2{}^{13}C^{18}OOH$ (bottom-left) and $NH_2CH_2{}^{13}C^{18}O^{18}OH$ (bottom-right), respectively. All masses corresponding to signals from glycine isotopologs, including 30 m/z, present a desorption peak at 245 K. The peak at 215 K corresponding to mass signal 78 in both bottom panels is due to a different species desorbing before glycine (see Supplementary Information).

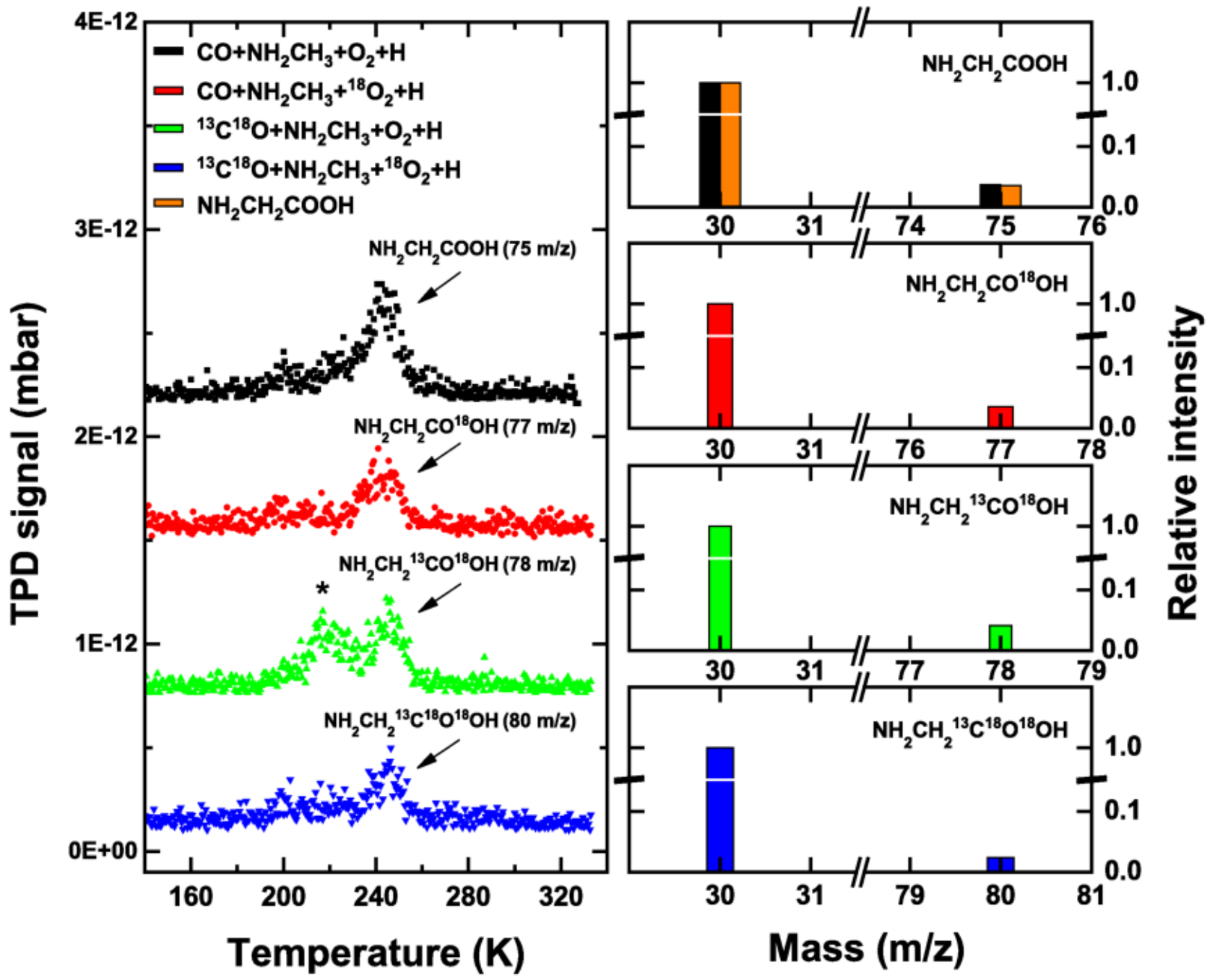

**Figure 3.** Quadrupole Mass Spectrometer-Temperature Programmed Desorption (QMS-TPD) data of non-fragmented glycine formed at 13 K and desorbed at 245 K. Left panel: Arrow indicates desorption peaks at 245 K for glycine and its isotopologs in our experiments 1-4 from Table 1. The symbol (*) indicates the desorption at 215 K of another species than glycine with mass 78 m/z, possibly formed during the TPD experiment (see Supplementary Information). Mass signals from experiments 1-4 highlight the expected mass shifts when isotopologs are used. TPD spectra are offset for clarity. In the right panels, selected mass-signals (30 and 75 m/z) from experiments 1 (black), 2 (red), 3 (green) and 4 (blue) from Table 1 are compared to the corresponding NIST mass-database values (orange).

Glycine formation is independently confirmed in the solid phase *in situ* before thermal desorption by means of RAIRS data. Figure 4 shows a RAIR difference spectrum (solid blue line) of two spectra acquired at 234 and 244 K, which highlights the residue sample of glycine obtained after desorption of all the more volatile ice components dominating the IR spectrum at lower temperatures. Transmission IR spectra of deposited pure glycine at 12 K and after heating to 240 K acquired during selected control experiments discussed in the Methods (dotted black and red lines, respectively) are presented for comparison. The RAIR difference spectrum presented here clearly show similarities with both transmission IR spectra indicating that in our main experiments glycine is firstly formed at low temperature and then experienced thermal heating. Glycine has been found in its different forms in deposited ice depending on the initial deposition temperature, thermal history and surrounding environment, e.g. surrounded by $H_2O$, $CO_2$ and $CH_4$[38]. Glycine can indeed act both as a proton acceptor, i.e. as a base, and proton donor, i.e. as an acid. Thus, depending on the surrounding conditions, four forms of glycine can be observed that are neutral ($NH_2CH_2COOH$), zwitterion ($H_3N^+CH_2COO^-$), cation ($H_3N^+CH_2COOH$) and anion ($NH_2CH_2COO^-$). To obtain the zwitterionic form from the neutral, glycine needs to be surrounded by other glycine molecules to donate/retrieve protons upon heating of the ice[39]. This happens for instance when the amine group deprotonates the carboxylic group forming a dipolar ion. In its cation and anion forms, the presence of other species is required, e.g. HCOOH or other acids and $NH_3$ or other bases, respectively. Therefore Figure 4 shows that pure glycine deposited at 20 K (dotted red line) is mostly glycine in its neutral form, while glycine deposited at 240 K (dotted black line) is mainly glycine in its zwitterionic form. The RAIR difference spectrum from experiment 1 of Table 1 (solid blue line in Figure 4) indicate that the neutral form is the dominant form of glycine in our surface formation experiments because of the presence of relatively strong absorption bands uniquely due to neutral glycine, for instance, at 1700 and 1230 cm$^{-1}$ (see symbols *). The IR absorption bands observed in the spectral range between 1800 and 1000 cm$^{-1}$ suggest that only a fraction of the glycine formed in our experiments is in its zwitterionic structure (see symbols ZG). The fact that we detect glycine in different forms in our main experiments is not surprising and it is an indication that glycine is firstly formed in its neutral form by means of surface reactions at low temperature (13 K) and then environmental conditions, i.e. ice composition and thermal heating to 230-240 K, determine its partial conversion to the zwitterion form. The non-fully conversion to its zwitterion suggests that most of the glycine molecules are formed in isolation within a water-rich ice matrix well representing dense cold molecular cloud conditions.

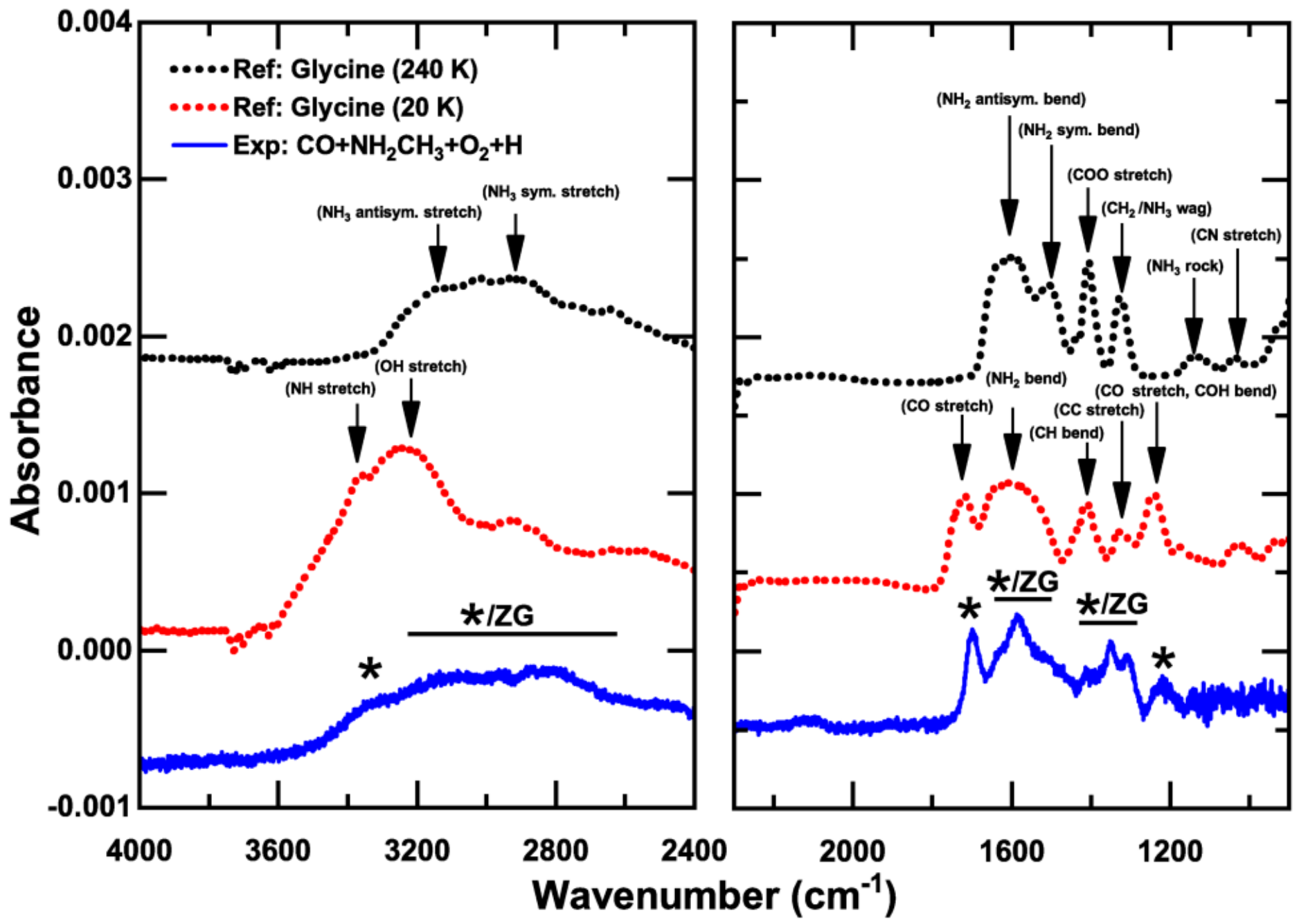

**Figure 4.** Reflection Absorption InfraRed (RAIR) data showing the presence of glycine ice in experiment 1 of Table 1. The two panels show different vibrational modes of glycine in the infrared spectral range. The solid blue line is the RAIR difference spectrum of two RAIR spectra recorded at 234 and 244 K highlighting the ice residue corresponding to glycine molecules as obtained from experiment 1 in Table 1. The dotted spectra are from additional control experiments of pure glycine deposited at 20 K (red) and heated to 240 K (black). The arrows indicate the main vibrational modes observed in the selected spectral ranges. The vibrational signatures in the red/black spectra show up in the blue spectrum. The residue spectrum of formed glycine (experiment 1) presents both neutral (*) and zwitterionic (ZG) features indicating that glycine is formed at low temperature before being thermally heated during the experiment[40-42]. Minor shifts (a few wavenumbers) in peak position between the glycine formation spectrum and the other two additional experiments are due to the fact that glycine formation was monitored in reflection, while pure glycine deposition was recorded in transmission in a different setup (see Methods).

Methylamine is not the only possible 'non-energetic' surface reaction intermediate to glycine (e.g. reaction $NH_2CH_2$ + HO-CO). We also tested under the same laboratory conditions another possible glycine surface formation reaction channel with acetic acid as a precursor (e.g. reaction $CH_2COOH$ + $NH_2$) and found the abstraction reaction of a hydrogen from the methyl group of acetic acid and from formic acid to be much less efficient than the analogue reaction for methylamine (see Fig. S3 and all control experiments in Table S1). Oba et al.[40] showed that H-D substitution occurs faster in the methyl group of methylamine than in the amino group. Therefore, we conclude that although the acetic acid channel should not be completely neglected, methylamine is a better precursor candidate towards glycine because its formation and destruction paths both lead to glycine in the solid phase at 10-20 K.

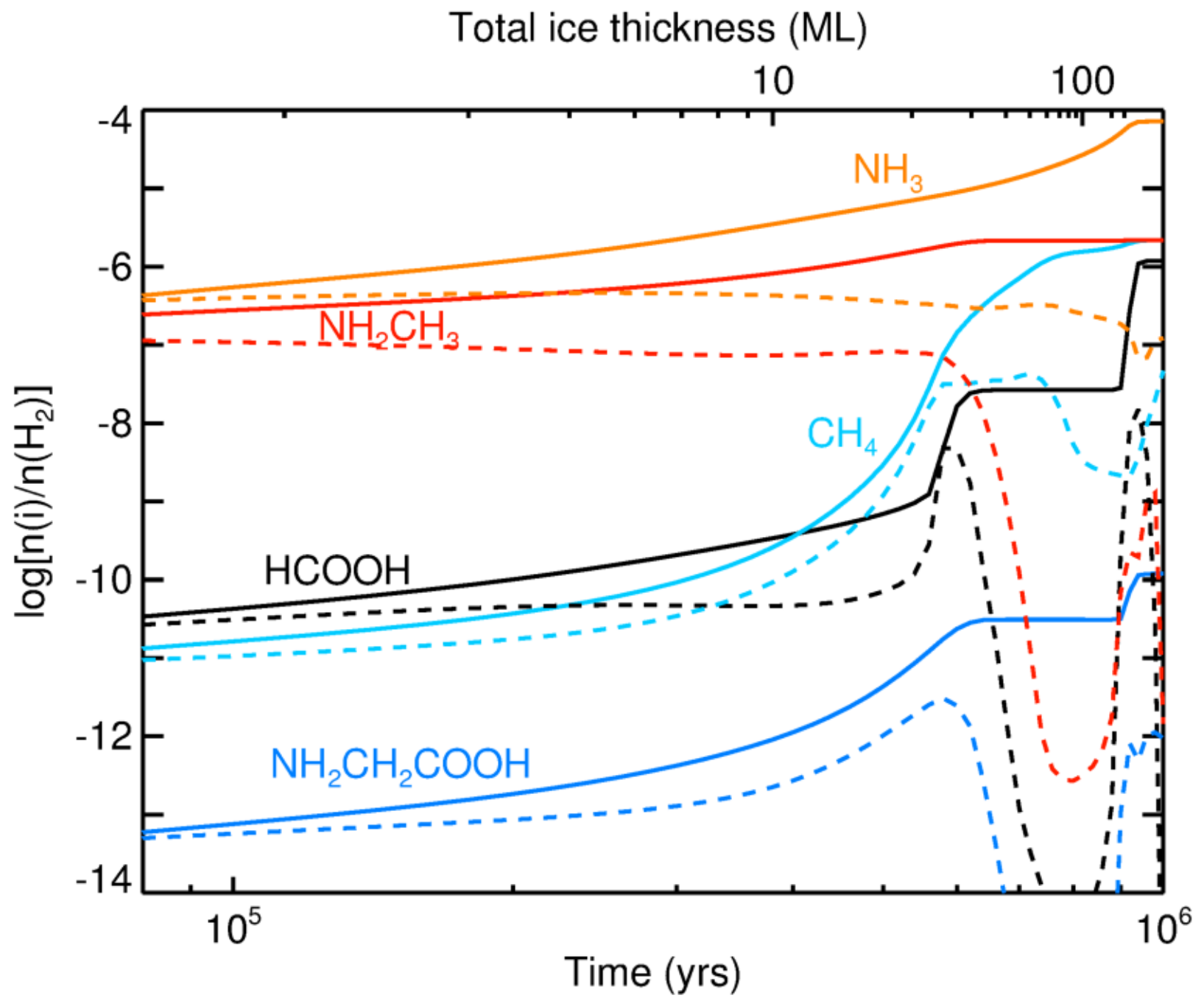

**Figure 5.** Abundances of solid species, including glycine and species involved in its surface formation, with respect to gas-phase $H_2$ during the collapse of a prestellar core, from Model 2. Solid lines indicate the total abundances of molecules in the dust-grain ice mantle; dashed lines of the same colors indicate the abundances present only in the chemically active upper layer of the ice. The final abundance of glycine in the ice mantle is around $10^{-10}$ with respect to $n_H$ at the end of the collapse over a period of 1 Myr.

The laboratory experiments described above show that a 'non-energetic' surface formation path for glycine at low temperatures is possible. Astrochemical models can extend on this, illustrating the importance of this reaction channel in the ISM by simulating the experimentally derived glycine formation channel in early phases of the star formation process. Two independent models are employed to this end: a microscopic kinetic Monte Carlo (kMC) model that treats the surface chemistry in detail and can be directly compared to the experiments and a rate equation model that can treat time-dependent physical conditions and can be more easily compared to astronomical observations. The kMC model (Model 1) keeps track of the positions of the adsorbed species. This is essential since at 10-20 K, apart from hydrogen atoms, the surface chemistry does not proceed through a diffusive mechanism but is based on a statistical probability that two reactants are formed in close proximity[41]. Rate equations (Model 2) can only account for this in an indirect way by changing the equations. They do however allow a full coupling with gas-phase chemistry. All models are described in more detail in the Supplementary Information. Model 1, based on Cuppen & Herbst[42] with site specific binding and diffusion[43], aims at reproducing experimental results by using a restricted reaction network. The model does not include any photochemical reactions. Input parameters are initial gas-phase abundance of species and desorption, diffusion and reaction rates. The included reaction routes and their corresponding rates have been verified experimentally or by quantum chemical calculations (see Table S2 and Fig. S4). It should be noted that at temperatures between 10-20 K diffusion and desorption rates are prohibitively low for all species other than hydrogen and the Model 1 is rather insensitive to these. Nevertheless, radical species can still form in proximity with each other and neighboring radicals with the correct geometric orientation can then recombine to form methylamine and glycine. This process does not require any UV irradiation to start the chemistry. Model 1 is then extended to ISM conditions with two different densities (see Fig. S5). Typical static dark-cloud models indicate that the H-atom abundance in the gas phase is around a few atoms per $cm^3$, the rest being in the form of $H_2$. Over time, the gas-phase abundances of the heavier atoms gradually drop as significant amounts of the overall O, C and N budgets are locked-up in the ices. A dark cloud hence goes from a low to high H-atom accretion rate relative to other atoms over time. Model 1 accounts for this effect. Results are shown in Table 2 and indicate that solid glycine can form in dense clouds through 'non-energetic' processing with a relative abundance to water of 0.04-0.07 %.

**Table 2.** Relative abundances of ice species in background sources and on comet 67P compared to corresponding values from Models 1 and 2.

| **Species** | **Bkg Stars** (ref.[28]) | **67P** (ref.[2,48]) Summer hem. | **Model 1** (*this work*) low $n_H$ [a] | **Model 1** (*this work*) high $n_H$ [b] | **Model 2** (*this work*) prestellar |
|---|---|---|---|---|---|
| $H_2O$ | 100 | 100 | 100 | 100 | 100 |
| CO | 9-67 | 2.7 | 8.0 | 17.3 | 19.7 |
| $CO_2$ | 14-43 | 2.5 | 4.0 | 6.9 | 9.3 |
| $CH_3OH$ | (< 1)-25 | 0.31 | 4.0 | 2.4 | 35.9 |
| $NH_3$ | < 7 | 0.06 | 8.6 | 5.6 | 21.3 |
| $CH_4$ | < 3 | 0.13 | 3.6 | 3.0 | 0.65 |
| HCOOH [c] | < 2 | 0.008 | 0.44 | 0.69 | 0.35 |
| $NH_2CH_3$ | | 0-0.16 | 2.9 | 1.8 | 0.65 |
| $NH_2CH_2COOH$ | <0.3 [d] | 0-0.16 | 0.04 | 0.07 | $3.5\times10^{-5}$ |

[a] Low density ($1\times10^4$ cm$^{-3}$). [b] High density ($2\times10^4$ cm$^{-3}$). [c] HCOOH may be a carrier of the 7.24 µm band. [d] Upper limit for the massive young stellar object W33 A taken from ref.[2].

A full gas-grain astrochemical kinetics model (*MAGICKAL*[15]; Model 2) is used to test the proposed production mechanisms for glycine under time-dependent physical conditions appropriate to a dense, prestellar core. Model 2 simulates the collapse of diffuse gas ($n_H = 3\times10^3$ cm$^{-3}$) under free-fall, reaching a peak total-hydrogen density of $2\times10^6$ cm$^{-3}$ after, approximately, 1 Myr. A chemical network with a total of 1366 chemical species, which include gas-phase, grain-surface and ice-mantle species, is used to simulate the chemistry over this period. The network is based on that of Garrod et al.[44] with the addition of all reactions and rates used in Model 1. Species are allowed to react in the top monolayer, with an ice mantle that builds up underneath as new material is deposited onto the surface. In order to demonstrate the ability of the tested mechanisms to operate without energetic processing, all chemical processing within the mantle phase is disabled in this model. Thus, the mantle composition is dependent solely on the time-dependent composition of the upper surface layer as new material is deposited on top. Most previous gas-grain models have typically considered only a purely diffusive mechanism for grain-surface reactions. Model 2 adapts the non-diffusive treatment introduced by Garrod & Pauly[45] for surface $CO_2$ production to all surface reactions, allowing the non-diffusive production of glycine and its precursors at low temperature; this scheme is presented in more detail by Jin & Garrod[46]. Figure 5 shows the fractional abundances with respect to molecular hydrogen of a selection of species, relevant in the formation of glycine as derived from the experimental work. Solid lines indicate the total abundances of molecules in the dust-grain ice mantle; dashed lines of the same colors indicate the abundances present only in the chemically active upper layer of the ice. Glycine is found to become abundant during the latter few hundred thousand years of evolution, when gas densities increase substantially and depletion of gas-phase material onto the grains becomes rapid. A final abundance with respect to $n_H$ of close to $10^{-10}$ is achieved within the ice ($3.5\times10^{-5}$ % with respect to water ice), purely through the new non-diffusive reaction of $NH_2CH_2$ with HO-CO. The production of $NH_2CH_2$ radicals has a substantial contribution from the reaction of CH with ammonia. As with $CO_2$, the production of HO-CO is dominated by non-diffusive reaction of CO and OH.

Our models, much like our extensive laboratory experiments, confirm that the link between methylamine and glycine found in comet 67P is expected to be present in the ISM in the solid phase. Our work indicates that glycine forms in a water-rich ice during the accretion of the first ice layers covering interstellar grains, i.e. before the 'catastrophic' freeze-out of CO molecules on dust grains. This finding puts the formation of glycine at a much earlier stage than previously thought. An early formation of glycine in the evolution of star-forming regions implies that glycine can be formed more ubiquitously in space and be preserved in the bulk of polar ices before inclusion in meteorites and comets during planet formation in protoplanetary disks surrounding newborn stars. Once formed, prestellar glycine can also become a precursor species to more complex molecules by 'energetic' and 'non-energetic' surface reaction routes. Moreover, the surface radical-radical recombination of $NH_2CH$ and HO-CO produces the precursor of all the proteinogenic α-amino acids, i.e. building blocks of the proteins relevant to life on Earth. Therefore simple proteinogenic α-amino acids such as alanine and serine are expected to be formed together with glycine in water-rich environments upon 'non-energetic' surface chemistry. Oba et al.[47] show that the 'non-energetic' deuterium (D) surface addition/abstraction reactions of glycine occur at the α-carbon under cold prestellar conditions. This suggests that surface substitution reactions of H atoms from the α-carbon with larger side chains (also known as R group that is, for instance, $CH_3$ for alanine and $CH_2OH$ for serine) can potentially lead to the formation of other more complex proteinogenic α-amino acids of biological relevance further enriching the organic molecular inventory in space that is later delivered to celestial bodies including planets like Earth. The conclusion is that glycine and possibly other building blocks of life are expected to be present at least in the solid phase across many star-forming environments including the coldest and earliest stages of solar-type systems formation. The so far lack of astronomical observations of glycine in the gas phase may be explained by the fact that glycine forms efficiently on ice grains in dark clouds and that its refractory nature allows it to remain in the solid phase longer than other species, possibly up to its inclusion into comets and planets.

## Acknowledgments

We thank T. Lamberts and I. Jiménez-Serra for stimulating discussions. This research was funded through a VICI grant of NWO (the Netherlands Organization for Scientific Research) and an A-ERC grant 291141 (CHEMPLAN). Financial support from the Danish National Research Foundation through the Center of Excellence "InterCat" (Grant agreement no.: DNRF150) and from NOVA (the Netherlands Research School for Astronomy) and the Royal Netherlands Academy of Arts and Sciences (KNAW) through a professor prize is acknowledged. S.I. acknowledges the Royal Society for financial support through the University Research Fellowship (UF130409), the University Research Fellowship Renewal 2019 (URF\R\191018) and the Research Fellows Enhancement Award (RGF\EA\180306), and the Holland Research School for Molecular Chemistry (HRSMC) for a travel grant. G.F. acknowledges the financial support from the European Union's Horizon 2020 research and innovation program under the Marie Skłodowska-Curie actions grant agreement n. 664931 and support from 'iALMA' grant (CUP C52I13000140001) approved by MIUR (Ministero dell'Istruzione, dell'Universitá e della Ricerca). A.R.C. and R.T.G. would like to thank NASA Astrophysics Research and Analysis Research program for funding through the grant NNX15AG07G. V.K. was funded by the NWO PEPSci (Planetary and ExoPlanetary Science) program. The described work advantaged from collaborations within the framework of the FP7 ITN LASSIE consortium (GA238258).

## Author contributions

S.I. initiated and managed the project. He wrote the manuscript with assistance from H.L., H.M.C., A.R.C., R.T.G., G.F. and K.-J.C., while E.F.v.D. linked the laboratory and modelling results to astronomical observations. S.I., K.-J.C., G.F., D.Q., and V.K. performed laboratory experiments. H.L. was responsible for the laboratory management. H.M.C., A.R.C., and R.T.G. developed and ran kinetic Monte Carlo simulations. M.J. and R.T.G. developed and ran gas-grain kinetics models. All authors contributed to data interpretation and commented on the paper.

## Methods

**Experimental.** Experiments are performed at the Laboratory for Astrophysics at Leiden Observatory using a suite of different experimental setups and techniques. The main experimental apparatus used here is SURFRESIDE[2], an ultra-high vacuum (UHV) setup designed to study 'non-energetic' atom addition and abstraction reactions in interstellar ice analogues under prestellar core conditions[1]. The base pressure of the main chamber is ~$10^{-10}$ mbar, i.e. the background $H_2O$ deposition rate (residual gas contamination) in the chamber is <$6\times10^{10}$ molecule $cm^{-2}$ $s^{-1}$. A rotatable gold-plated copper substrate is placed at the center of the main chamber and cooled by a closed-cycle helium cryostat. A resistive heating wire allows to vary the substrate temperature between 8 and 450 K. Two silicon diodes are used to monitor the temperature with 0.5 K absolute accuracy. Gases are introduced in the main chamber and deposited with monolayer precision onto the substrate surface through two metal deposition lines mounted under an angle of 22 degrees from the substrate surface normal. Only high purity chemicals and some selected isotopologs are used in all the experiments, i.e. $NH_2CH_3$ (>98 %), $^{16}O_2$ (99.999 %), $^{18}O_2$ (97 %), $^{12}CO$ (99.99 %), $^{13}C^{18}O$ (95 %), HCOOH (>98 %), $CH_3COOH$ (99.7 %) and glycine (99.9 %). In control experiments where formaldehyde is deposited (see Supplementary Information), paraformaldehyde (95 %) powder is connected to the pre-pumped dosing line and is thermally decomposed by a bath of hot water in order to form $H_2CO$ vapor. All ices are grown with sub-monolayer precision with deposition rates provided in Table 1 and S1, where main experiments and controls are listed, respectively. All deposition rates and times are chosen to keep the thickness of the ices comparable to the thickness of interstellar ice mantles on dust grains (~10-100 ML). Moreover, deposition rates of different species are selected according to practical reasons: the deposition rate of $O_2$ is the lowest to minimize the amount of non-hydrogenated $O_2$ trapped in the ice, to increase the production of OH radicals and to minimize the formation of $H_2O$ and $H_2O_2$ in the ice; H-atom deposition rate is the highest to ensure that hydrogenation reactions lead to the formation of new stable species and radicals in the ice; other ratios are based on optimized experimental settings for the production of glycine, allowing for its detection by means of infrared spectroscopy and mass spectrometry.

Two atom sources are connected to the main chamber through shutters. Each source has an angle of 45 degrees with respect to the substrate. H/D-atoms are introduced using a Hydrogen Atom Beam Source (HABS)[2] mounted in a UHV atom line. A second UHV atom line hosts a Microwave Atom Source (MWAS)[3] that can be used to dissociate species like $NH_3$ in its fragments. In each atom line, a nose-shape quartz pipe is placed along the path to efficiently quench and thermalize excited atoms, radicals and non-dissociated molecules through multiple collisions with the walls of the pipe. Fluxes used in the experiments 1-6 are listed in Table 1. In our experiments, hydroxyl radicals are formed in the ice to ensure that they are promptly thermalized before further reaction. It should be noted that experiments are here set to have the lowest number of involved initial reactants in the ice to fully control and characterize reaction schemes. Moreover, our experiments have the unique potential to study selected parts of the reaction network at interstellar relevant conditions; i.e. under the very same ice conditions and processes found in space, namely the presence of radicals in a water-rich ice at 10-20 K, with multiple reactions competing with each other. Consequently, our experiments do not have to necessarily start from interstellar relevant ices, but ultimately simulate interstellar relevant surface reaction routes.

The ice composition is monitored *in situ* by means of reflection absorption infrared (RAIR) spectroscopy in the range between 4000-700 $cm^{-1}$ with a spectral resolution of 1 $cm^{-1}$ using a Fourier transform infrared (FTIR) spectrometer. The main chamber gas-phase composition is monitored by a quadrupole mass spectrometer (QMS), which is placed behind the rotatable substrate, and is used during temperature programmed desorption (TPD) experiments, in which the ice is heated at a constant rate (Table 1) until its complete sublimation. Spectra of pure deposited glycine ice are obtained using a separate multi-functional, high-vacuum (HV) ice setup[4]. This system is equipped with a sublimation oven, which allows for the preparation of selected films of non-volatile species, which can then be studied *in situ* by means of Fourier transform infrared spectroscopy in transmission mode. A commercial sample of glycine is used without further purification. Glycine is heated to a temperature of 388 K in a small oven positioned close to the substrate holder, where it is deposited at 12 K and then heated to different temperatures. Spectra of pure glycine ice are shown in Figure 4 and are compared to the results obtained with SURFRESIDE[2], experiment 1 in Table 1.

**Kinetic Monte Carlo (kMC) model.** Model 1 is based on Cuppen & Herbst[5] with site specific binding and diffusion[6]. The model uses a flat geometry of the surface with periodic boundary conditions. Binding energies are taken from the recent collection in Penteado et al.[7]. The H-atom diffusion rate is site dependent and it covers the lower end of the rates reported in refs.[8,9]. The program includes some exothermicity that allows products to hop a few sites after reaction. The gas phase starts with a fixed initial composition which is depleted in all species except for H and $H_2$ self-consistently during the simulation accounting for the accreted species. No gas-phase chemistry is considered. The considered reaction network contains 32 species, 48 surface reactions and no photodissociation reactions. The network is discussed in more detail in Supplementary Information and Table S2.

**Gas-grain chemical kinetics model.** Model 2 (*MAGICKAL*) is a rate-based astrochemical kinetics code, that uses modified reaction rates[10] to approximate stochastic behaviour on the grain surfaces where appropriate[11]. A uniform ratio of diffusion barriers to surface binding energies of 0.35 is assumed. The physical collapse of the gas from density $3\times10^3$ cm$^{-3}$ to $2\times10^6$ cm$^{-3}$ takes $9.267\times10^5$ years; the physical conditions are held steady beyond this time until the end-time of 1 million years is reached. The dust temperature falls from ~16 to 8 K during the collapse, following the treatment of Garrod & Pauly[12]. The gas temperature is held at a constant 10 K throughout. The visual extinction ranges from 2 to ~150 mag during the collapse. Gas-phase and grain-surface photo-dissociation is active throughout the model run (in tandem with chemical reactions), with the grain-surface rates set to a fraction one third[13] of the gas-phase rates, using the same branching ratios. No photo-dissociation (or other chemical/physical process) is considered within the ice mantle, only on surfaces. The initial elemental abundances are those used by Garrod[11] and others. A fraction 0.002 of hydrogen is initially in atomic form, with the remainder starting as $H_2$; this value is an approximate steady-state value under the initial physical conditions for $A_V = 2$ mag.

The treatment for non-diffusive reactions is based on that presented by Garrod & Pauly[12]; these authors introduced a non-diffusive mechanism for $CO_2$ production, through the formation of OH (via diffusive H and O addition) in the presence of CO, leading to immediate reaction to give $CO_2$ + H. The similar treatment adopted here considers the prior chemical production rates and surface coverages of each of the reactants in a particular subsequent reaction. Thus, an initial round of rate calculations is completed before the *follow-on* or *three-body reaction* rates are calculated. The total rate of each follow-on reaction is the sum of two terms, each of which is equal to the production rate of one reactant multiplied by the surface coverage of the other, the latter of which constitutes an approximation to the encounter probability per production event. In cases where there is an activation energy barrier, the follow-on reaction rate is multiplied by an efficiency that takes into account the competition between reaction and surface diffusion of either species, in line with regular diffusive treatments. In order to allow all possible follow-on reactions leading to glycine, three rounds of such reactions are allowed, with production rates from each round used to calculate subsequent production rates, with diminishing importance after each round. The chemical network used[11] includes three main mechanisms for glycine production, including addition of $NH_2CH_2$ and HO-CO radicals, the addition of $NH_2$ and $CH_2COOH$ radicals, and addition of $NH_2CH_2CO$ and OH radicals. The main mechanism of glycine production in the models is the first of these processes in agreement with the interpretation of the experimental results presented in this work.

## Supplementary Information

Other candidate molecules for mass signal 75 m/z desorption peak at 245 K

Glycine is not the only candidate species for mass signal 75 m/z in the QMS-TPD experiment $NH_2CH_3$+CO+$O_2$+H. Hydrogenation experiments at low temperature and thermal processing of the ice during TPD can potentially lead to the formation of other COMs in the ice that present fragmentation mass signal 75 m/z such as acetohydroxamic acid ($CH_3CONHOH$), methyl carbamate ($NH_2COCH_3O$), methylcarbamic acid ($CH_3NHCOOH$) and glycolamide ($NH_2COCH_2OH$). Acetohydroxamic acid isotopologs present mass shifts different from glycine. Therefore Figure 2 of the main text indicates that desorption of acetohydroxamic acid at 245 K can be excluded in our main experiments. One of the main fragmentation mass signals for methyl carbamate, methylcarbamic acid and glycolamide is mass 44 m/z, i.e. the relative intensity of mass signal 44 m/z is orders of magnitude higher than that of mass 75 m/z for all the aforementioned molecules[1,2], in contrast with the case of glycine for which the relative intensities for masses 44 and 75 m/z are comparable. Therefore, since in our experiments 1-4 of Table 1 mass signals 44, 45 and 46 m/z do not present clear intense desorption peaks at 245 K on top of their background signal, we can conclude that methyl carbamate, methylcarbamic acid and glycolamide are not the species responsible for the mass 75 m/z peak at 245 K (see Fig. S1).

Figure S1 shows a selection of mass signals of species desorbing from the ice in the temperature range 150-330 K during the QMS-TPD of the main $NH_2CH_3$+CO+$O_2$+H experiment (experiment 1 in Table 1). Among all masses monitored, masses 30 and 75 m/z allow for the identification of glycine during its desorption at 245 K. Masses 44, 45 and 46 m/z are the main signals for non-fragmented $CO_2$ and possibly fragmented methyl carbamate, methylcarbamic acid and glycolamide desorbing at 215-230 K (also see inset in Figs. S1 and S2). Figure S1 clearly shows that there is nearly no desorption peak for masses 44, 45 and 46 m/z at 245 K. A small change in slope of mass signal 44 m/z can be seen in the inset of Fig. S1 at 245 K corresponding to a bump for mass signal 30 m/z. The change in slope of mass 44 m/z corresponds to the desorption of an amount of molecules quantitatively comparable to the desorption of mass signal 75 m/z at the same temperature. The latter finding is in agreement with the NIST data for the fragmentation of glycine supporting its identification in our experiments by means of QMS-TPD. Further experiments discussed in the main text involving isotopolog species also confirm this result. Finally, the infrared fingerprints of methyl carbamate, methylcarbamic acid and glycolamide do not reproduce the infrared profile of the residue of experiment 1 of Table 1 shown in Figure 4 of the main text[1,3,4]. For instance, the residue spectrum of experiment 1 due to glycine does not present the intense, sharp absorption band at ~ 3370 $cm^{-1}$ of methylcarbamic acid ice[1]. This is a further indication that solid glycine is the best candidate to be formed in the ice and then to desorb around 245 K.

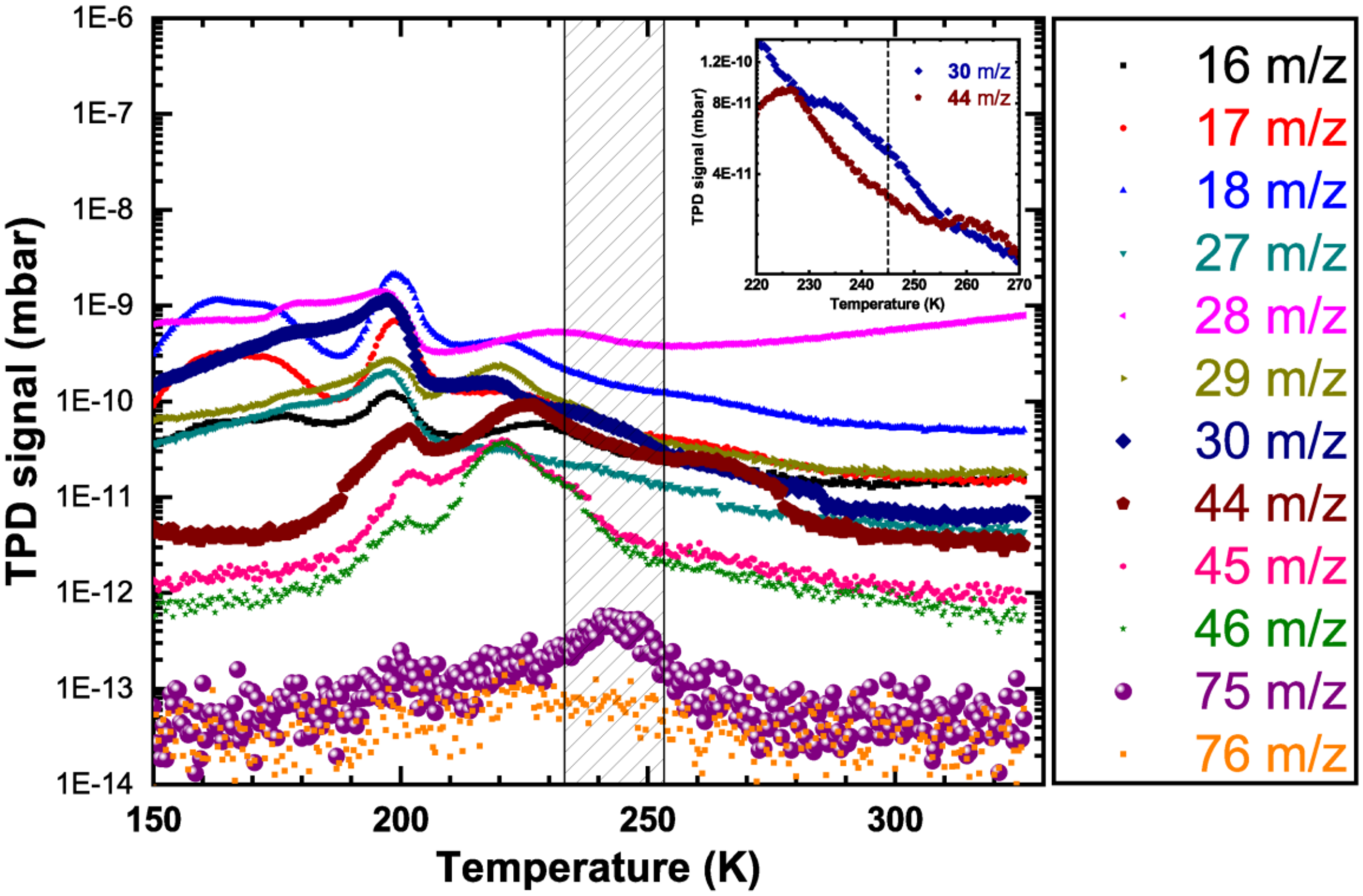

**Fig. S1.** Selection of mass signals of species desorbing from the ice in the temperature range 150-330 K during the QMS-TPD of the main $NH_2CH_3$+CO+$O_2$+H experiment (Exp. 1). The desorption peak at 245 K of masses 30 and 75 m/z is due to glycine. Inset shows mass signal 30 and 44, where a peak at 225 K is only visible for mass 44 m/z and a peak at 245 K only for mass 30 m/z.

Control experiments on thermally activated surface reactions

A long list of control experiments is performed to further constrain the formation of glycine in the ice. Control experiments include deposition of pure ices and co-deposition (i.e. simultaneous deposition of molecules, radicals and atoms) experiments all performed at 13 K as shown in Table S1. As discussed in the main text, glycine formation is carried-out by co-depositing methylamine, carbon monoxide, molecular oxygen and hydrogen atoms to enhance its formation and detection by means of RAIRS and QMS-TPD. In this work, systematic control experiments are performed to test that glycine is formed at low temperature and before heating the ice during a TPD experiment (see, for instance, comparison of spectra in Figure 4 of main text). Other control experiments are designed to investigate thermally activated surface reactions (see Fig. S2) showing that glycine is not one of the surface products when methylamine and carbon dioxide, methylamine and formaldehyde or methylamine and formic acid are, respectively, co-deposited at 13 K and then heated up to 350 K.

**Table S1.** List of the most relevant control experiments.

| Control Experiments Thermal processing | T K | Flux ($CO_2$) $10^{12}$ mol. cm$^{-2}$ s$^{-1}$ | Flux (HCOOH) $10^{12}$ mol. cm$^{-2}$ s$^{-1}$ | Flux ($H_2CO$) $10^{12}$ mol. cm$^{-2}$ s$^{-1}$ | Flux ($NH_2CH_3$) $10^{12}$ mol. cm$^{-2}$ s$^{-1}$ | time min | TPD rate K/min | |
|---|---|---|---|---|---|---|---|---|
| $CO_2$:$NH_2CH_3$=1:10 | 13 | 0.6 | - | - | 6.6 | 60 | 5 | |
| HCOOH:$NH_2CH_3$=1:10 | 13 | - | 0.5 | - | 6.6 | 60 | 5 | |
| $H_2CO$:$NH_2CH_3$=1:10 | 13 | - | - | 0.7 | 6.6 | 60 | 5 | |
| $H_2CO$:$NH_2CH_3$=1:1 | 13 | - | - | 6.7 | 6.6 | 60 | 5 | |
| **Other Control Experiments** | **T** K | **Flux (HCOOH)** $10^{12}$ mol. cm$^{-2}$ s$^{-1}$ | **Flux ($CH_3COOH$)** $10^{12}$ mol. cm$^{-2}$ s$^{-1}$ | **Flux ($NH_2CH_3$)** $10^{12}$ mol. cm$^{-2}$ s$^{-1}$ | **Flux (D)** $10^{12}$ mol. cm$^{-2}$ s$^{-1}$ | **Flux ($O_2$)** $10^{12}$ mol. cm$^{-2}$ s$^{-1}$ | **time** min | **H** [a] **Abs.** |
| HCOOH | 13 | 6.7 | - | - | - | - | 35 | - |
| HCOOH+D | 13 | 6.7 | - | - | 5 | - | 120 | N |
| HCOOH+D+$O_2$ | 13 | 0.2 | - | - | 5 | 1.4 | 60 | N |
| HCOOH+D+O | 13 | 0.2 | - | - | 5 | 1.4×*d* [b] | 60 | N |
| $CH_3COOH$ | 13 | - | 3.3 | - | - | - | 35 | - |
| $CH_3COOH$+D | 13 | - | 3.3 | - | 5 | - | 120 | N |
| $CH_3COOH$+D+$O_2$ | 13 | - | 0.3 | - | 5 | 1.4 | 120 | N |
| $CH_3COOH$+D+$O_2$ | 13 | - | 0.3 | - | 5 | 0.1 | 60 | N |
| $CH_3COOH$+D+O | 13 | - | 1.4 | - | 5 | 0.1×*d* [b] | 60 | N |
| $NH_2CH_3$ | 13 | - | - | 0.8 | - | - | 35 | - |
| $NH_2CH_3$+D | 13 | - | - | 0.6 | 5 | - | 80 | Y |
| $NH_2CH_3$+D+$O_2$ | 13 | - | - | 0.6 | 5 | 1.4 | 90 | Y |
| $NH_2CH_3$+D+O | 13 | - | - | 0.6 | 5 | 1.4×*d* [b] | 90 | Y |

[a] Evidence for hydrogen abstraction from parental molecule reacting with H or OH. [b] Dissociation coeff. = 8-12 %.

Figure S2 shows that heating our HCOOH:$NH_2CH_3$ mixtures does not produce species with mass 75 m/z desorbing in the range 150-330 K. On the other hand, thermally processed mixtures of $H_2CO$:$NH_2CH_3$ and $CO_2$:$NH_2CH_3$ do show the desorption of species with mass signal 75 m/z at 215 K and 226 K, respectively. However, we exclude this to be glycine because of the low desorption temperature and because the fragmentation ratio of masses 30, 44, and 75 m/z does not correspond to the one for glycine as published by NIST mass-database and as observed in our experiment 1 of Table 1 at 245 K. In fact, in all the mentioned control experiments on thermal processing mass 44 m/z is more than one order of magnitude more intense than mass 75 m/z. Thus, our QMS-TPD experiments show that the mass signal 75 m/z found at 245 K in the main experiment 1 of Table 1 is due to glycine, confirming its formation upon 'non-energetic' surface processing of the ice. The unambiguous identification of species formed upon thermal heating during control experiments is beyond the scope of this work. Nevertheless, we note that the desorption temperature for mass signals 44-46 and 75 m/z at 215 K and 226 K in the $H_2CO$:$NH_2CH_3$ and $CO_2$:$NH_2CH_3$ experiments, respectively, is in agreement with the desorption temperature for some peaks at 210-230 K due to mass signals 44-46 m/z in experiment 1 of Table 1 shown in Figure S1. Mass signal 78 m/z in experiments 3 and 4 of Table 1 presents a distinct peak at 215 K (see Figures 2 and 3 of the main text). Therefore, it is likely that, apart from glycine desorbing at 245 K, some other COMs are formed during our main experiments (Exp. 1-4 of Table 1) and desorb prior to glycine sublimation.

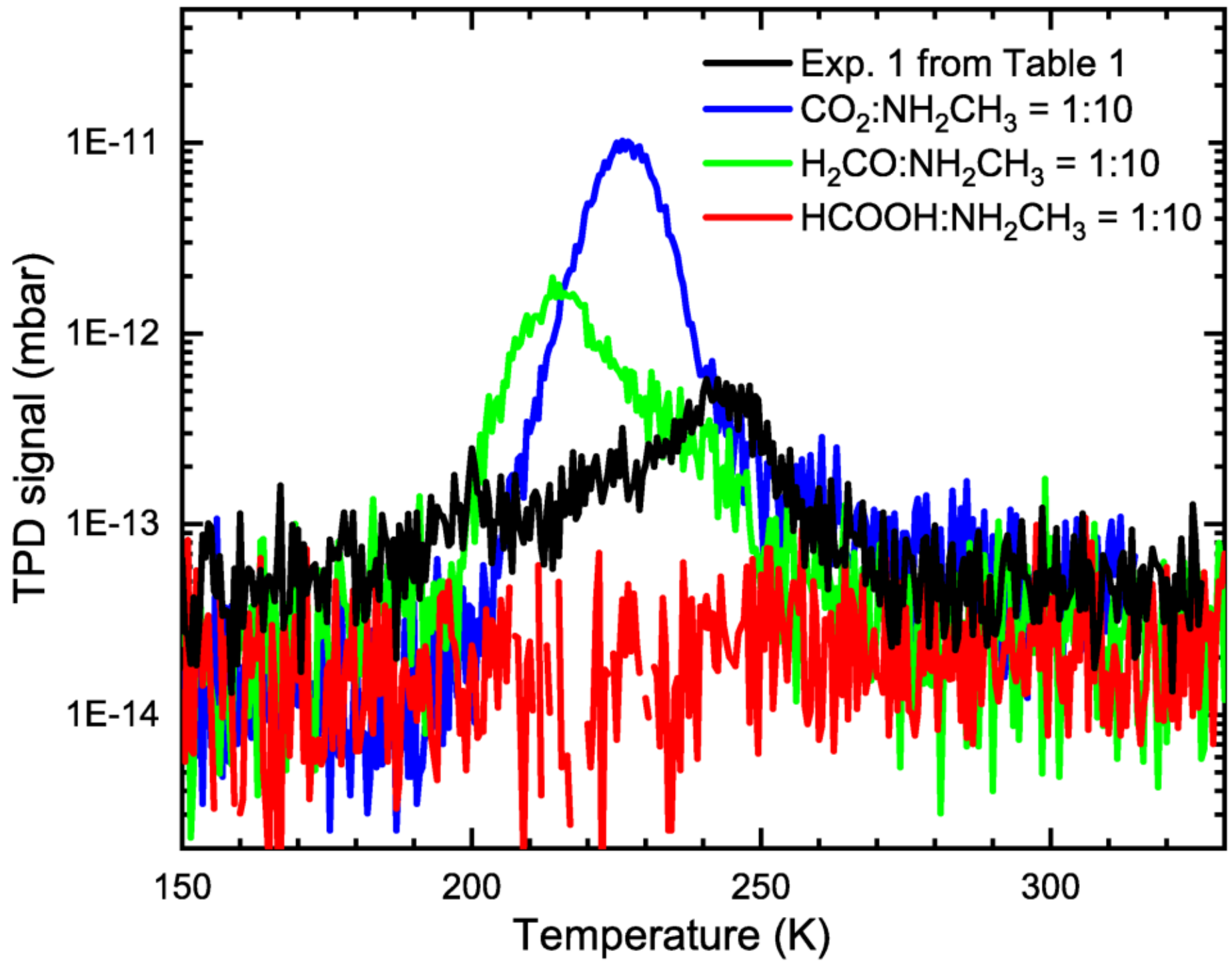

**Fig. S2.** Mass signal 75 m/z from desorbing species in the temperature range 150-330 K during the QMS-TPD of the main $NH_2CH_3$+CO+$O_2$+H experiment (black – Exp. 1 of Table 1). Mass signal 75 m/z from experiment 1 is compared with the same mass signal from mixtures $CO_2$:$NH_2CH_3$ = 1:10 (blue), $H_2CO$:$NH_2CH_3$ = 1:10 (green) and HCOOH:$NH_2CH_3$ = 1:10 (red). Relative intensities of peaks from control experiments should not be compared to the intensity of the peak for experiment 1 because of the different experimental conditions of experiment 1 compare to the other controls.

Desorption temperature of formed glycine

It is worth noting that desorption of a multilayer deposited pure zwitterionic glycine ice occurs above 300 K[5]. Our QMS-TPD data show that intact neutral glycine (m/z = 75 amu) formed in our experiments desorbs at 245 K (right panel of Figure 2). Esmaili et al.[5] show that under their experimental conditions neutral glycine formed upon low energy electron irradiation of a $CO_2$:$CH_4$:$NH_3$ ice mixture desorbs at 275 K. Therefore, while its neutral form desorbs at lower temperatures, other forms of glycine are more refractory. The authors also highlight that the glycine desorption temperature is highly sensitive to its structure, e.g. neutral vs zwitterionic form, and therefore depends on the surrounding environment and the thickness of the ice, e.g. (sub)monolayer vs multilayer[5]. Thus, differences in desorption temperature between our work and ref.[5] are most likely due to different environmental conditions, specifically, ice composition and thickness.

Control experiments on abstraction surface reactions

Other control experiments that can potentially lead to the formation of glycine precursors are listed at the end of Table S1. Such experiments aim at investigating the hydrogen and hydroxyl radical abstraction reactions from formic acid to form the HO-CO complex, from acetic acid to form the $CH_2COOH$ radical and from methylamine to form the $NH_2CH_2$ radical, respectively. Figure S3 shows that after careful comparison of the three H-abstraction reactions, the methylamine channel is the most efficient one. Results are also discussed in the main text.

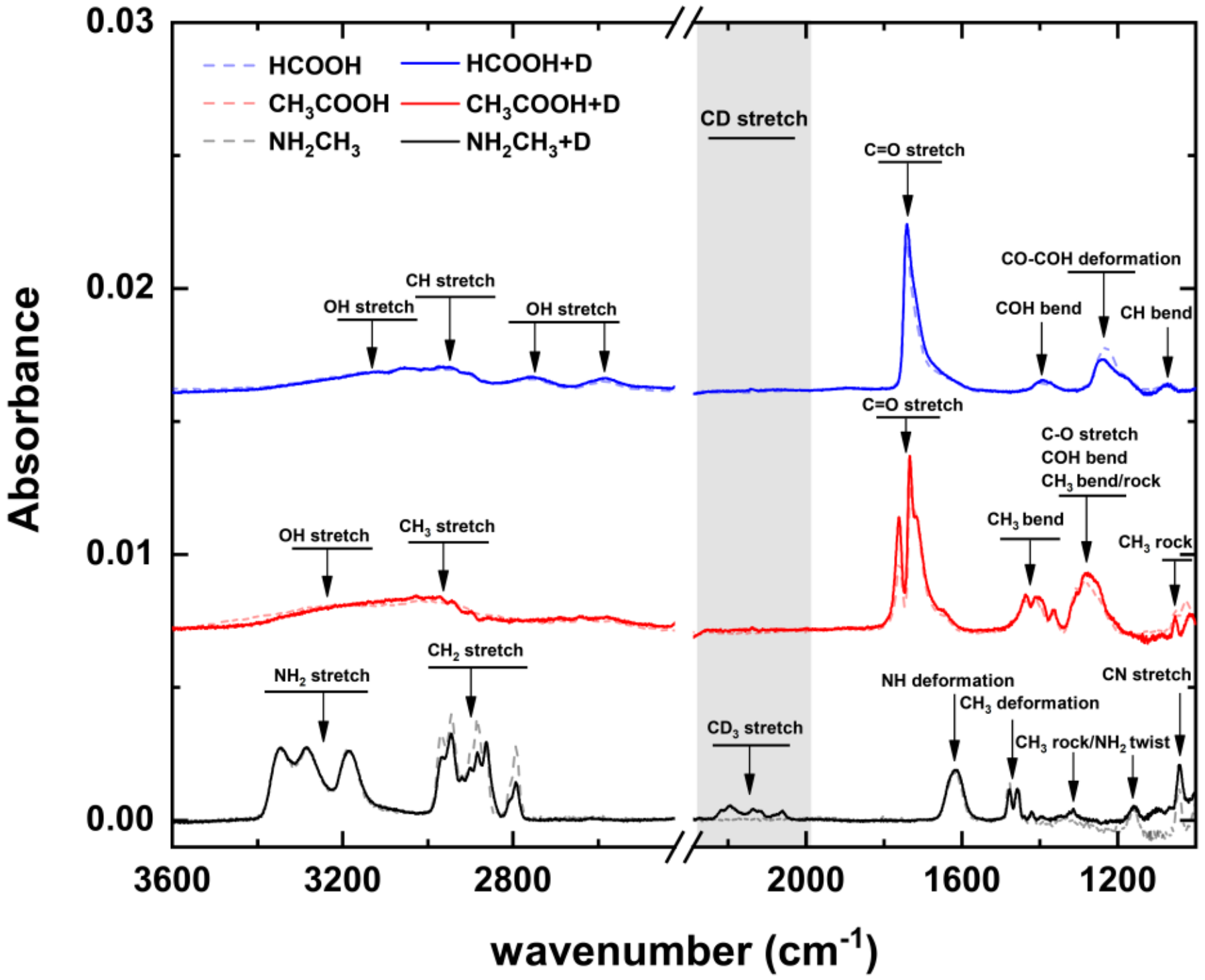

**Fig. S3.** Solid methylamine (black line), acetic acid (red line) and formic acid (blue line) exposed to deuterium atoms at 13 K. Spectra of pure methylamine, acetic acid and formic acid are shown for comparison (dashed lines). Grey area shows the CD stretch spectral region of deuterated methylamine. More details are available in Table S1. Hydrogen abstraction followed by deuterium addition reactions on the methyl functional group is observed only in the case of $NH_2CH_3$ + D. Similar results, not shown here, are obtained when OD radicals are used instead of D atoms. In the latter case, also deuterated water ice is observed. This finding confirms that reactions 25 and 26 from Table S2 occur under interstellar analogue conditions and methylamine is a possible precursor of glycine in space. Spectra are normalized and offset for clarity. Assignments are made according to selected references (HCOOH)[6], ($CH_3COOH$)[7,8], and ($NH_2CH_3$)[9].

**Table S2.** Reaction network as included in Model 1.

| | Reaction | k ($s^{-1}$) | Branching ratio | Refs |
|---|---|---|---|---|
| 1 | $H + H \rightarrow H_2$ | $2 \times 10^{11}$ [a] | | |
| 2 | $H + O \rightarrow OH$ | $2 \times 10^{11}$ | | |
| 3 | $H + OH \rightarrow H_2O$ | $2 \times 10^{11}$ | | |
| 4 | $CO + H \rightarrow HCO$ | $2 \times 10^{-3}$ | | 10 |
| 5 | $HCO + H \rightarrow H_2CO$ | $2 \times 10^{11}$ | 0.33 | |
| 6 | $HCO + H \rightarrow H_2 + CO$ | | 0.67 | |
| 7 | $H_2CO + H \rightarrow HCO + H_2$ | $2 \times 10^{-4}$ | 0.5 | 10,11 |
| 8 | $H_2CO + H \rightarrow H_3CO$ | | 0.5 | 10,11 |
| 9 | $H_3CO + H \rightarrow H_3COH$ | $2 \times 10^{11}$ | | |
| 10 | $CO + OH \rightarrow HO\text{-}CO$ | $7 \times 10^{-2}$ | 0.5 | |
| 11 | $CO + OH \rightarrow CO_2 + H$ | | 0.5 | |
| 12 | $HO\text{-}CO + H \rightarrow CO_2 + H_2$ | $2 \times 10^{11}$ | 0.5 | 12-14 |
| 13 | $HO\text{-}CO + H \rightarrow HCOOH$ | | 0.5 | |
| 14 | $N + H \rightarrow NH$ | $2 \times 10^{11}$ | | |
| 15 | $NH + H \rightarrow NH_2$ | $2 \times 10^{11}$ | | |
| 16 | $NH_2 + H \rightarrow NH_3$ | $2 \times 10^{11}$ | | |
| 17 | $C + H \rightarrow CH$ | $2 \times 10^{11}$ | | |
| 18 | $CH + H \rightarrow CH_2$ | $2 \times 10^{11}$ | | |
| 19 | $CH_2 + H \rightarrow CH_3$ | $2 \times 10^{11}$ | | |
| 20 | $CH_3 + H \rightarrow CH_4$ | $2 \times 10^{11}$ | | |
| 21 | $CH_4 + OH \rightarrow CH_3 + H_2O$ | $5 \times 10^{2}$ | | 15 |
| 22 | $NH_2 + CH_3 \rightarrow NH_2CH_3$ | $2 \times 10^{11}$ | | |
| 23 | $NH_3 + CH \rightarrow NCH_4$ | $2 \times 10^{11}$ | | 16 |
| 24 | $NCH_4 + H \rightarrow NH_2CH_3$ | $2 \times 10^{11}$ | | |
| 25 | $NH_2CH_3 + H \rightarrow NCH_4 + H_2$ | $9 \times 10^{-1}$ | | 9 |
| 26 | $NH_2CH_3 + OH \rightarrow NCH_4 + H_2O$ | $4 \times 10^{-3}$ | | |
| 27 | $NCH_4 + HO\text{-}CO \rightarrow NH_2CH_2COOH$ | $8 \times 10^{-2}$ | | |
| 28 | $OH + H_2 \rightarrow H_2O + H$ | $2 \times 10^{5}$ | | 17 |
| 29 | $O + O \rightarrow O_2$ | $2 \times 10^{11}$ | | |
| 30 | $O_2 + H \rightarrow HO_2$ | $1 \times 10^{11}$ | | 18 |
| 31 | $H + HO_2 \rightarrow OH + OH$ | $2 \times 10^{11}$ | 0.94 | 19 |
| 32 | $H + HO_2 \rightarrow H_2 + O_2$ | | 0.02 | |
| 33 | $H + HO_2 \rightarrow H_2O + O$ | | 0.05 | |
| 34 | $OH + OH \rightarrow H_2O_2$ | $2 \times 10^{11}$ | 0.87 | |
| 35 | $OH + OH \rightarrow H_2O + O$ | | 0.13 | |
| 36 | $H_2O_2 + H \rightarrow H_2O + OH$ | $3 \times 10^{4}$ | | 20 |
| 37 | $N + N \rightarrow N_2$ | $2 \times 10^{11}$ | | |
| 38 | $N + O \rightarrow NO$ | $2 \times 10^{11}$ | | |
| 39 | $NO + H \rightarrow HNO$ | $2 \times 10^{11}$ | | |
| 40 | $HNO + H \rightarrow H_2NO$ | $2 \times 10^{11}$ | 0.5 | |
| 41 | $HNO + H \rightarrow NO + H_2$ | | 0.5 | |
| 42 | $HNO + O \rightarrow NO + OH$ | $2 \times 10^{11}$ | | |
| 43 | $O + NH \rightarrow HNO$ | $2 \times 10^{11}$ | | |
| 44 | $N + NH \rightarrow N_2 + H$ | $2 \times 10^{11}$ | | |
| 45 | $NH + NH \rightarrow N_2 + H_2$ | $2 \times 10^{11}$ | | |
| 46 | $C + O \rightarrow CO$ | $2 \times 10^{11}$ | | |
| 47 | $CH_3 + OH \rightarrow CH_3OH$ | $2 \times 10^{11}$ | | |

[a] All reactions with $k = 2 \times 10^{11}$ are considered barrierless.

Reaction network in kMC Model 1

Model 1 firstly aims to reproduce the experimental finding (Fig. S4) and subsequently extends the proposed glycine formation scheme to the ISM (Fig. S5). The reactions included in the network of Model 1 are given in Table S2 with literature references for the adopted rates. The network includes two routes to form methylamine: through reaction 22, and reactions 23 and 24, respectively. Reaction 22 is the most likely route under the experimental conditions. Reaction 23 is based on quantum chemical calculations by Blitz et al.[21] and involves a reaction between a radical and an abundant stable species. Under interstellar conditions this is more likely to occur according to statistical probability than, for instance, $NH_2$ + $CH_3$, which is a reaction between two transient species. Reaction 23 can result in the formation of $NHCH_3$, which can rearrange to the more stable $NH_2CH_2$ (Ref.[16]). Both species should easily hydrogenate to methylamine. For this reason, $NCH_4$ is used in the reaction network as a collective term for both isomers.

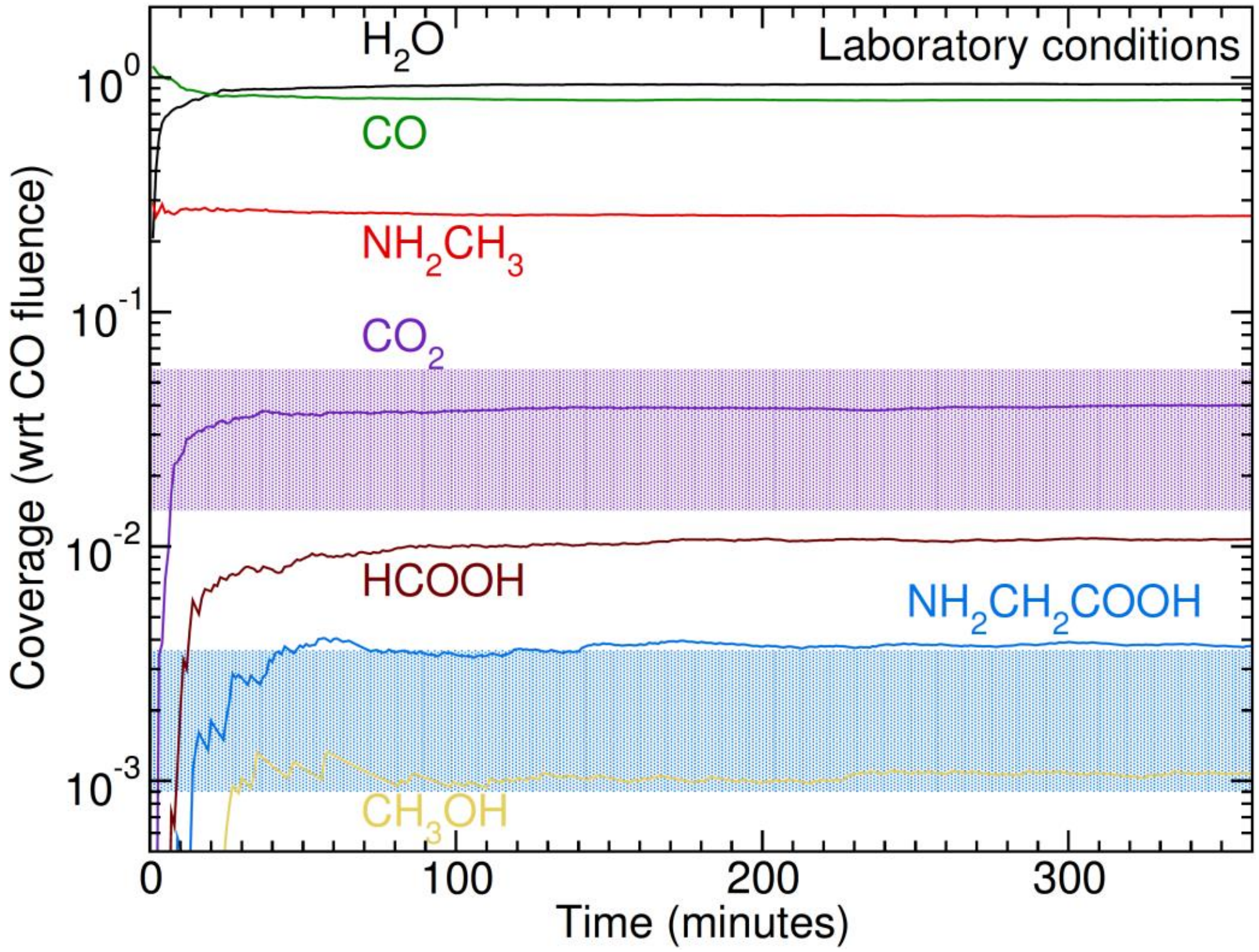

**Fig. S4.** Simulation of experimental conditions using Model 1. $NH_2CH_3$+CO+(H, $H_2$)+$O_2$ with respect to deposited CO simulating experiment 1 starting from a bare surface. The colored stripes ($CO_2$ dark pink and glycine light blue) indicate experimental measurements assuming an uncertainty of a factor of 2. Temperature is 14 K for all the simulations.

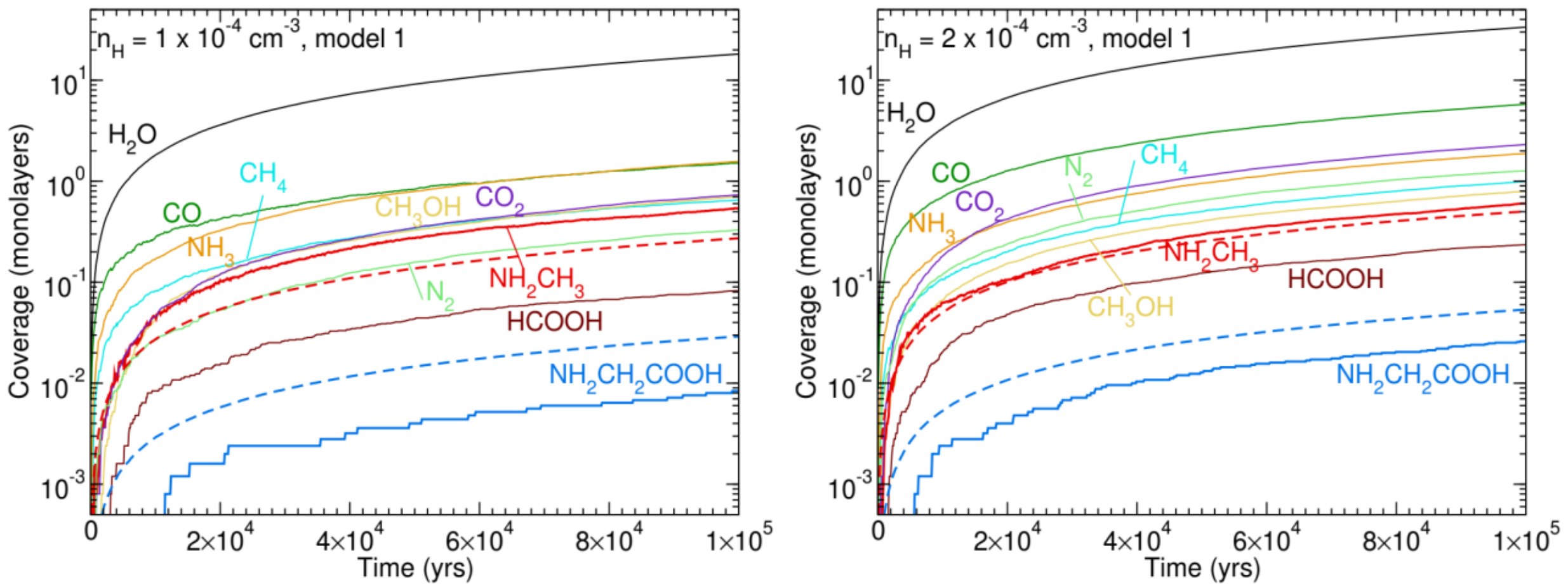

**Fig. S5.** Simulations under ISM conditions at 10 K for Model 1. Abundances are plotted in number of monolayers, where 1 monolayer corresponds roughly to $10^{15}$ molecules $cm^{-2}$. The corresponding final abundances with respect to $H_2O$ ice are reported in Table 2. Left panel shows data for the low-density condition ($n_H=1\times10^4$ $cm^{-3}$); right panel for the high density ($n_H=2\times10^4$ $cm^{-3}$). Dashed lines indicate the observed abundances for methylamine and glycine found in comet 67P w.r.t. $H_2O$ ice. The lower production of glycine w.r.t. observation values can be explained by the fact that Model 1 is conservative and does not include any energetic process that can alter the amount of glycine in the ice. It is likely that Comet 67P has indeed been to some extent exposed to solar wind and cosmic ray irradiation throughout its lifetime in the Solar System.

One of the main assumptions is the lifetime of the HO-CO complex in the ice. HO-CO is observed to be stable at detectable levels in a polar ice[12]. It can subsequently react with atomic hydrogen with three exit channels[13]. HO-CO could also directly dissociate to $CO_2$ + H through tunnelling[14] like in the gas phase route, which is added in the model by a direct route of CO and OH leading to $CO_2$. The laboratory $CO_2$/CO ratio is used here to constrain reactions 10 and 11. The resulting rate is in the range of quoted barriers in the quantum chemistry literature (for an overview see Ref.[22]). Abstraction from methylamine by OH is also not well constrained due to the formation of an activated complex[23,24]. Abstraction by H probably proceeds through tunnelling. Unfortunately, only barrier values are available[9]. Reaction 46 is included to avoid an overproduction of $CH_4$ and $O_2$ in the ice. The CO formed in the ice through reaction 46 does not significantly affect the outcome on the formation of glycine and methylamine.

Fluxes in kMC Model 1

Experimentally the deposition of $NH_3$ dissociation products results in an $N_2$-rich ice rather than $NH_3$-rich. To reproduce the laboratory results, the $NH_3$ fragments composition needs to be N-rich. Here $NH_3$:$NH_2$:NH:N=5:5:5:85 is assumed in Model 1. The H-atom flux is calculated accordingly, where 50 % of the H atoms is assumed to combine to $H_2$. Both H and $H_2$ are assumed to stick with a 30 % efficiency at 14 K in accordance with literature values[25]. This leads to fluxes relative to the total $NH_3$ flux of ($NH_3$+$NH_2$+NH+N):H:$H_2$=100:40:20. This is a rather low atom flux compared to previous experiments of, for instance, hydrogenation of CO[10-12], where typically H-flux/CO-flux = 10 are applied. Consequently, we expect hydrogen atoms to predominantly react with the oxygen and nitrogen bearing species, since most of these reactions occur barrierless and to form very little methanol, which involves tunnelling through several barriers. Table S3 compares the steady state abundances for the experimental simulation from Model 1 with the values from the laboratory experiments showing a good agreement within laboratory and simulation uncertainties.

**Table S3.** Relative steady state values for the experimental simulations of Model 1 vs the corresponding laboratory experiment values (experiments 1-6 in Table 1). Band strength values for CO, $CO_2$ and glycine are from refs.[26,27].

| | $CO_2$/CO (%) | Gly/CO (%) |
|---|---|---|
| **Exp 1** | 6.1 | 0.16 |
| **Exp 2** | 2.1 | 0.16 |
| **Exp 3** | 3.8 | 0.22 |
| **Exp 4** | 4.8 | 0.23 |
| **Exp 5** | 5.8 | 0.14 |
| **Exp 6** | 4.9 | 0.14 |
| **Model 1** | 4.0 | 0.38 |

For Model 1, initial gas-phase abundances taken for the ISM simulations are given in Table S4. Relative abundances are obtained by a gas-grain model of a translucent cloud ($n_H$ = 500 cm$^{-3}$ and $A_V$ = 2 mag) after $10^6$ years. These abundances, other than for H atoms, are then used as starting conditions for two molecular cloud models with densities of low $n_H = 1\times10^4$ cm$^{-3}$ and high $n_H = 2\times10^4$ cm$^{-3}$ (Fig. S5). For Model 1, the gas-phase abundance of $H_2$ is reduced to 100 cm$^{-3}$ to minimise simulation times while keeping $H_2$ the most dominant species. Short simulations of higher $H_2$ abundance did not lead to significantly different results.

**Table S4.** Gas-phase abundances to which the grain is exposed as used in Model 1.

| **Species** | **Model 1** ***n*(X) (cm$^{-3}$)** (low $n_H$) | ***n*(X) (cm$^{-3}$)** (high $n_H$) |
|---|---|---|
| H | 3 | 3 |
| $H_2$ | 5,000 | 10,000 |
| O | 2.3 | 4.5 |
| CO | 0.85 | 1.7 |
| N | 0.39 | 0.77 |
| C | 0.18 | 0.36 |

Figure S5 shows the abundances of a selection of molecules in the low and high $n_H$ models over a period of $10^5$ years. The abundances of most species become stable after several $10^4$ years. Between the low and high $n_H$ models, we find a somewhat different chemistry. In the high $n_H$ model, the lower H-atom accretion rate w.r.t. the other species leads to less saturated species, such as $CH_4$ and $NH_3$, and significantly increases the effectiveness of the formation pathway to glycine discussed throughout this paper. The abundances of selected species from Model 1 are shown in Table 2 of the main article. In Model 1, glycine abundances were found to be around 0.04% and 0.07% with respect to $H_2O$ by the end of each run, with a greater abundance produced under the higher relative hydrogen accretion rate, i.e. in the low $n_H$ model. The formation of glycine relies on the formation of HO-CO and $NH_2CH_2$. With the increased amount of heavier radical species like N and C in the high $n_H$ model, the relatively mobile hydrogen atoms are less likely to react with each other than to find an alternative reaction partner, meaning that more $CH_2$ and $NH_2$ are formed (precursors to $NH_2CH_2$). Accordingly, HO-CO is less likely to react with H than $NH_2CH_2$ to produce glycine, due to the higher abundance of $NH_2CH_2$. The difference in glycine abundances between the two H-atom density settings (low and high $n_H$) highlights the regimes between conditions, where there are always more radicals present than hydrogen atoms (high $n_H$), and where there are always H atoms in excess on the surface (low $n_H$).